\documentstyle[twocolumn,aps,pra,epsfig]{revtex}
\newcommand{\beq}{\begin{equation}}
\newcommand{\eeq}{\end{equation}}
\newcommand{\eq}[1]{Eq. (\ref{#1})}

\def\opra{Phys.~Rev.~A.~}

\def\prl#1#2#3{Phys.~Rev.~Lett.~{\bf #1},\ #2\ (#3)}

\def\pra#1#2#3{Phys.~Rev.~A~{\bf #1},\ #2\ (#3)}

\def\a{{\hat a}}
\def\ad{{\hat a}^\dagger}
\def\b{{\hat b}}
\def\bd{{\hat b}^\dagger}
\def\na{{\hat n}_a}
\def\lx{{\hat{\cal L}}_x}
\def\ly{{\hat{\cal L}}_y}
\def\lz{{\hat{\cal L}}_z}
\def\ddt{\frac{d}{dt}}
\def\ddtau{\frac{d}{d\tau}}
\def\sx{s_x}
\def\sy{s_y}
\def\sz{s_z}
\def\lj{{\hat{\cal L}}_j}
\def\li{{\hat{\cal L}}_i}

\def\H{\hat H}

\begin{document}
\title{Quantum effects on the dynamics of a two-mode atom-molecule 
 Bose-Einstein condensate}
\author{A. Vardi$^1$, V. A. Yurovsky$^{1,2}$, and J. R. Anglin$^3$}
\address{$^1$ITAMP, Harvard-Smithsonian Center for Astrophysics, 
60~Garden~Street, Cambridge MA 02138}
\address{$^2$School of Chemistry, Tel Aviv University, 69978
Tel Aviv, Israel}
\address{$^3$Center for Ultracold Atoms, MIT 26-237, 
77 Massachusetts Avenue, Cambridge MA 02139}
\maketitle

\begin{abstract}
We study the system of coupled atomic and molecular condensates within the 
two-mode model and beyond mean-field theory (MFT). Large amplitude 
atom-molecule coherent oscillations are shown to be damped by the 
rapid growth of fluctuations near the dynamically unstable molecular 
mode. This result contradicts earlier predictions about the recovery 
of atom-molecule oscillations in the two-mode limit. The frequency of
the damped oscillation is also shown to scale as $\sqrt{N}/\log N$ with the total
number of atoms $N$, rather than the expected pure $\sqrt{N}$ scaling. 
Using a linearized model, we obtain 
analytical expressions for the initial depletion of the molecular 
condensate in the vicinity of the instability, and show that the 
important effect neglected by mean field theory is an initially non-exponential
`spontaneous' dissociation into the atomic vacuum. Starting with a small 
population in the atomic mode, the initial dissociation rate is sensitive 
to the exact atomic amplitudes, with the fastest (super-exponential) rate 
observed for the entangled state, formed by spontaneous dissociation.
\end{abstract}
~\\

Recent photoassociation \cite{WFHRH00} and Feshbach 
resonance \cite{IASMSK98,JILA} experiments suggest the 
possibility of producing molecular Bose-Einstein condensates (BEC)  
\cite{javanainen,timmermans,heinzen,YBJW99,AV99,WJ00}.
Large amplitude coherent oscillations between an atomic 
BEC and a molecular BEC have been theoretically predicted 
\cite{javanainen,timmermans,heinzen}. A common
theme to these studies is the use of the Gross-Pitaevskii (GP) mean-field
theory (MFT), reducing the full multi-body problem into a set of 
two coupled nonlinear Schr\"odinger equations. These are then solved 
numerically to obtain the Josephson-type dynamics of the coupled 
atomic and molecular fields. 

The simple GP dynamics is substantially affected by condensate
depletion due to inelastic collisions \cite{timmermans,YBJW99,YBJW00}, 
spontaneous emission, and the inclusion of non-condensate 
modes \cite{YBJW00,MJT99,holland,goral,hope}. Consequently, the
atom-molecule oscillations are expected to be strongly damped 
under current experimental conditions. The proposed remedy 
for this detrimental effect involves a recovery of an effective
two-mode dynamics \cite{goral}, thereby preventing the buildup
of thermal population. 

In this article we point out that even in the perfect two-mode 
limit, MFT fails to provide long-term predictions due to strong 
interparticle entanglement near the dynamically unstable molecular 
mode. Quantum corrections to MFT appear in
the vicinity of its dynamical instabilities, on time scales that
grow only logarithmically with the number $N$ of condensate 
particles \cite{castin,vardi,anglin}. Thus, even in the absence
of a 'proper' thermal bath of non-condensate states, the mean-field 
equations are coupled to a reservoir of Bogoliubov fluctuations
\cite{vardi,habib}. The rapid growth of these fluctuations near
the instability is analogous to the rapid population of the thermal 
cloud, similarly inhibiting the mean-field atom-molecule 
oscillations. Our results, obtained using the numerical solution
of exact quantum equations, go beyond the Hartree-Fock-Bogoliubov 
approach \cite{holland}. The leading quantum effect is identified 
as a non-exponential spontaneous decay of the molecular condensate 
and the dynamics is shown to be highly sensitive to the initial 
conditions. We note that similar quantum corrections have been 
predicted for parametric oscillations 
in quantum optics \cite{michael}.
        
We consider the simplest model of the atom-molecule 
condensate, in which particles can only populate two second-quantized
modes: an atomic mode, associated with the creation and
annihilation operators $\a$ and $\ad$ and a molecular 
mode, associated with the creation and annihilation 
operators $\b$ and $\bd$. The two modes are coupled 
by means of a near-resonant two-photon transition or a Feshbach 
resonance, with a coupling frequency $\Omega$ and detuning $\Delta$. 
 Setting the zero energy to the energy 
of the molecular mode the two-mode Hamiltonian reads
\beq
\H=\frac{\hbar\Delta}{2}\ad\a+
\frac{\hbar\Omega}{2}\left(\ad\ad\b+\bd\a\a\right)~.
\label{hamiltonian}
\eeq
For $\Delta=0$, the Hamiltonian of \eq{hamiltonian} is 
identical \cite{javanainen} 
to the well known Hamiltonian describing the optical process
of parametric oscillations \cite{yariv}, where  
dissociation is equivalent to parametric
downconversion and association is the analog of second-harmonic 
generation. We will
take $\Omega$ to be real and positive without loss of generality, 
since the relative phase between the modes is determined 
up to an additive  constant and the overall sign of $H$ is 
insignificant.

We obtain a generalization of the Bloch representation 
for the two-mode system (similarly to the approach taken in 
Refs. \cite{vardi,anglin}), by introducing the three operators,
\begin{eqnarray}
\label{lxyz}
\lx&\equiv&\sqrt{2}\frac{\ad\ad\b+\bd\a\a}{N^{3/2}}\nonumber\\
\ly&\equiv&\sqrt{2}\frac{\ad\ad\b-\bd\a\a}{iN^{3/2}}\\\
\lz&\equiv&\frac{2\bd\b-\ad\a}{N}~,\nonumber
\end{eqnarray}
where $N$ denotes the total number of atoms.
Discarding c-number terms, the Hamiltonian of \eq{hamiltonian} 
then takes the simple form
\beq
\H=\hbar\left[\left(\frac{N}{2}\right)^{3/2}\Omega\lx
-\frac{N}{4}\Delta\lz\right]~,
\eeq
and the Heisenberg equations of motion for the three 
operators of \eq{lxyz} read
\begin{eqnarray}
\label{qeom}
\ddt\lx&=-&\Delta\ly\nonumber\\
\ddt\ly&=&-\frac{3}{4}\sqrt{2N}\Omega
\left[\left(\lz-1\right)\left(\lz+\frac{1}{3}\right)\right]+\Delta\lx
+\sqrt{2\over N}\Omega\nonumber\\
\ddt\lz&=&-\sqrt{2N}\Omega\ly~.
\end{eqnarray}
These three operators do {\it not} represent SU(2); but all three commute with
the conserved total atom number 
$N=\hat{a}^\dagger\hat{a}+2\hat{b}^\dagger\hat{b}$.

The mean field approximation is invoked by approximating second 
order expectation values $\langle\li\lj\rangle$
as products of the first order moments $\langle\li\rangle$ 
and $\langle\lj\rangle$ \cite{vardi,anglin}
\beq
\langle\li\lj\rangle\approx\langle\li\rangle\langle\lj\rangle~.
\label{mfa}
\eeq
Approximation (\ref{mfa}) is correct to ${\cal O}(1/\sqrt{N})$.
Thus we can also neglect the $c$-number term $\sqrt{2/N}\Omega$ 
in \eq{qeom}.  
Defining $\vec{\bf s}\equiv(\langle\lx\rangle,\langle\ly\rangle,
\langle\lz\rangle)$, rescaling the time as $\tau=\sqrt{N}\Omega t$
and using \eq{mfa}, we obtain the mean-field equations
\begin{eqnarray}
\label{mfe}
\ddtau\sx&=&-\delta\sy\nonumber\\
\ddtau\sy&=&-\frac{3\sqrt{2}}{4}
\left(\sz-1\right)\left(\sz+\frac{1}{3}\right)+\delta\sx\\
\ddtau\sz&=&-\sqrt{2}\sy~,\nonumber
\end{eqnarray}
where the dimensionless rescaled detuning $\delta$ is defined 
as $\delta\equiv\Delta/\sqrt{N}\Omega$ (Eqs. (\ref{mfe}) 
are equivalent to Eqs. (32) of Ref. \cite{YBJW00} without the 
inelastic collision terms).  
Lyapunov analysis of Eqs. (\ref{mfe}) shows that as long as 
$|\delta|<\sqrt{2}$ the stationary point $\vec{\bf s}=(0,0,1)$ 
corresponding to the entire population being in the molecular mode, 
is dynamically unstable because any small perturbation to the mean-field 
equations (\ref{mfe}) would trigger the parametric oscillation. 
In the vicinity of this point, MFT is expected to break down on a time
scale which is only logarithmic in $N$ \cite{castin,vardi,anglin}. 
In order to verify this prediction, we solve the full N-body 
problem by fixing $N$, thereby restricting the available 
phase-space to Fock states of the type $|n,(N-n)/2\rangle$ 
with $n$ atoms and $(N-n)/2$ molecules, 
where $n=0,2,4,...,N$. Thus we obtain an $N/2+1$ dimensional 
representation for the Hamiltonian and the $N$-body density 
operator $\hat\rho$. The quantum solution is then obtained 
by numerically propagating $\hat\rho$ according to the  Liouville 
von-Neumann equation
\beq
i\hbar\dot{\hat\rho}=[\H,\dot{\hat\rho}]~,
\eeq
and the expectation values of the three operators of \eq{lxyz} 
are retrieved as $s_i=Tr(\rho\li)$.

In what follows, we shall assume that $\delta=0$, as required 
to obtain unity amplitude atom-molecule oscillations.
In Fig. 1 we plot the expectation value $\sz$, corresponding to 
the population difference between the modes, as a function of the 
rescaled time $\tau$, for various values of the total particle 
number. The initial conditions are $\vec{\bf s}=(0,0,-1)$, 
corresponding to an initially populated atomic mode. The mean
field solution (dotted line) depicts the convergence of $\sz$  
to the unstable fixed point. The quantum solutions  
(identical to the results of Ref. \cite{javanainen} obtained 
by solving the $N$-particle Schr\"odinger equation), initially 
follow the mean-field evolution closely. However, in the vicinity 
of the molecular mode ($\sz=1$) the quantum trajectories break 
away from the mean-field prediction on a timescale that grows
only logarithmically with $N$. Thus, the oscillation frequency 
scales with $\sqrt{N}/\log(N)$ as opposed to the expected scaling
with $\sqrt{N}$. Moreover, the oscillations are damped
by the strong entanglement near the molecular mode, in 
complete analogy with the damping of oscillations when the 
two-mode system is coupled to external thermal modes. Full
Rabi-type coherent oscillations can only be observed for a 
single pair of atoms.

We note that the results of Fig. 1 can not be reproduced by a 
Hartree-Fock-Bogoliubov approach \cite{holland}. 
In order to obtain the damping of coherent oscillations, 
one has to go  deeper in the Bogoliubov-Born-Green-Kirkwood-Yvon (BBGKY) 
hierarchy of equations of motion, and maintain a number of equations 
comparable to the total number of particles $N$.
   
\begin{figure}
\begin{center}
\epsfig{file=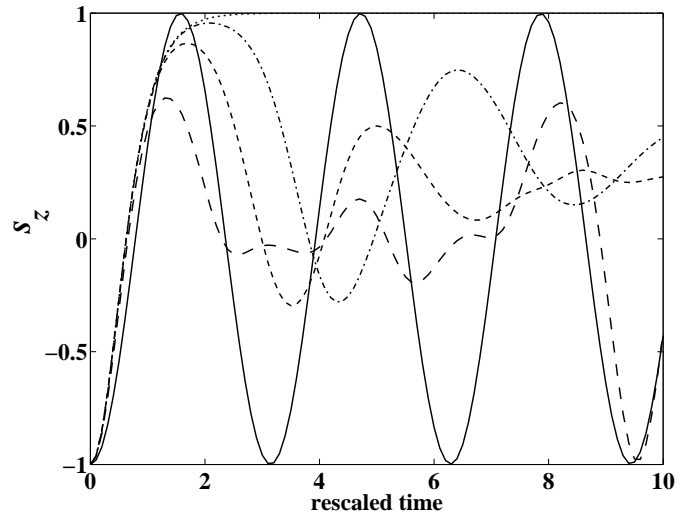,width=0.8\columnwidth,angle=90}
\end{center}
\caption{Numerically calculated population imbalance between atoms 
and molecules, as a function of the rescaled time $\tau$ for 2 (-----), 
10 ($---$), 100 (- - -), and 1000 ($-\cdot-$) particles, compared 
with the mean-field prediction ($\cdots$).}
\label{f1}
\end{figure}

Equations (\ref{qeom}) are equivalent to the equations of 
motion for the two annihilation operators $\a$ and $\b$,
\begin{mathletters}
\begin{eqnarray}
i\dot{\a}&&=\frac{\Delta}{2}\a+\Omega\b\ad 
\label{adot}
\\
i\dot{\b}&&=\frac{\Omega}{2}\a\a~.
\label{bdot}
\end{eqnarray}
\end{mathletters}
To appreciate why MFT fails as it does, we will now focus our
attention on the vicinity of the dynamically unstable all-molecule state, 
using a linearized model in 
which the molecular annihilation operator $\b$ is replaced 
by a c-number $b$ of ${\cal O}(\sqrt{N/2})$. In this approximation, 
which is valid as long as the population of the molecular state is 
large and the effect of its depletion on the atomic population 
growth rate can be neglected, \eq{adot} becomes
\beq
i\dot{\a}=\frac{\Delta}{2}\a+\Omega b\ad~. 
\label{adotp}
\eeq
Equation (\ref{adotp}) in combination with its complex conjugate provide an 
autonomic set of two linear operator equations, which can be solved
using common methods reducing it to an eigenproblem. When the molecular 
mode is dynamically unstable ($|\delta|<\sqrt{2}$) we have
\begin{equation}
\Omega |b| > |\Delta|/2~,  \label{cond_gain}
\end{equation} 
and the exact solution of Eq.\ (\ref{adotp}) takes the rapidly growing form 
\begin{equation}
\hat{a}(t)=\hat{a}(0)\cosh\lambda t
- {i \over \lambda}\left[\frac{\Delta}{2}\hat{a}(0)
+\Omega b \hat{a}^\dag(0)\right] \sinh\lambda t~, 
\label{asol}
\end{equation}
where
\begin{equation}
\lambda=\sqrt{\Omega^2|b|^2-(\Delta/2)^2}~.
\end{equation}
The time-dependence of the atom number operator 
$\na(t)=\ad(t)\a(t)$ is thus given as
\beq
\langle \na(t)\rangle=n_{\text{sp}}(t)+n_{\text{st}}(t)~, 
\label{nt}
\eeq  
where the term 
\beq
n_{\text{sp}}(t)={\Omega^2|b|^2 \over \lambda^2}\sinh^2\lambda t
\label{nsp}
\eeq
not accounted for by MFT, corresponds to spontaneous dissociation into 
the atomic vacuum and the term 
\begin{eqnarray}
n_{\text{st}}(t)=\langle \na(0)\rangle\left(\cosh 2\lambda t
+\frac{\Delta^2}{2\lambda^2}\sinh^2\lambda t \right) \nonumber
\\ 
-{\Omega \over \lambda} \text{Im}
\biggl[\langle \hat{a}(0) \hat{a}(0)\rangle b^*\left(\sinh 2\lambda t 
-i \frac{\Delta}{\lambda}\sinh^2\lambda t \right)
\biggr] \label{nst}
\end{eqnarray}
depicts stimulated dissociation taking place when the atomic state 
is initially populated. The two terms on the r.h.s of \eq{nst} 
correspond to non-coherent and coherent initial occupation, 
respectively.

Starting from the dynamical instability ($|b|=\sqrt{N/2}$) with zero 
atomic occupation, and assuming zero detuning, the initial evolution 
of the expectation value $\sz(t)$ is given according to Eqs 
(\ref{nt})-(\ref{nst}) by the purely spontaneous process,
\beq
\sz(t)=1-{2 \over N}\sinh^2 (\tau/\sqrt{2})~,
\label{szq}
\eeq
According to \eq{szq} the initial decay of the atomic mode 
is non-exponential as the leading decay term is quadratic 
rather than linear in $t$. This behavior is in accordance 
with the initial non-exponential decay of a general spontaneous 
emission process \cite{bzp}. The quadratic gain of the atomic 
population is confirmed by the results of Fig. 2 where we compare the 
initial depletion of the molecular mode according to \eq{szq} 
with exact quantum results obtained for various values of $N$. 
The agreement is initially excellent 
until the occupation of the atomic mode becomes significant 
compared to the molecular occupation. Moreover, it is evident from
\eq{szq} and confirmed by the results of Fig. 2, that the time at 
which the quantum spontaneous emission term will become significant, 
grows only logarithmically with $N$, in agreement with our 
prediction.  
\begin{figure}
\begin{center}
\epsfig{file=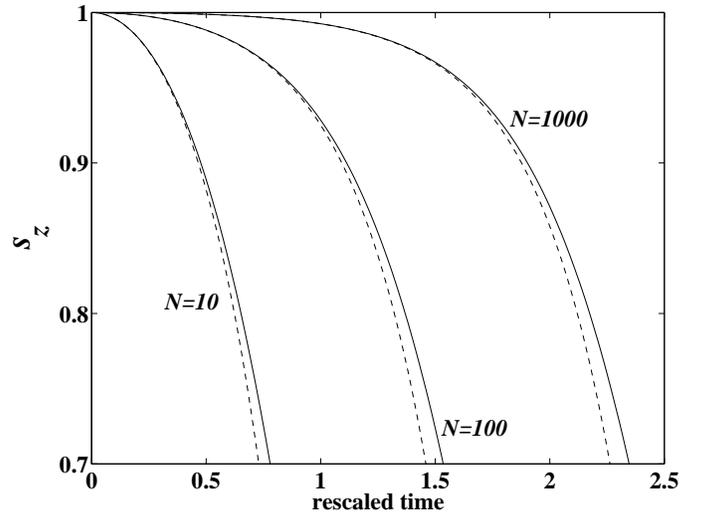,width=0.8\columnwidth,angle=90}
\end{center}
\caption{Depletion of the purely molecular mode for $N=10$, 
100  and 1000 particles, according to \eq{szq} ($- - -$), 
compared with 'exact' numerical results (-----).}
\label{f2}
\end{figure}

Equation (\ref{nsp}) was already obtained in Ref.\ \cite{goral}, 
as the asymptotic expression at $t\rightarrow \infty$. 
However, at this limit, the atomic occupation becomes comparable 
to the molecular population, 
and the depletion of the molecular mode should be taken into 
account. Nevertheless, \eq{nsp} is an exact solution to the model
of \eq{adotp}, applicable at small $t$. 

It is interesting to note that when $|\delta|>\sqrt{2}$ the molecular 
mode is stabilized. Consequently, the exact solution of \eq{adotp} when 
\begin{equation}
\Omega |b| < |\Delta|/2  \label{cond_osc}
\end{equation}
is an oscillatory function of the form
\begin{equation}
\hat{a}(t)=\hat{a}(0)\cos|\lambda| t
- {i \over |\lambda|}\left
[\frac{\Delta}{2}\hat{a}(0)+\Omega b \hat{a}^\dag(0)\right] 
\sin|\lambda| t 
\label{aoscsol}
\end{equation}
where now
\begin{equation}
|\lambda|=\sqrt{(\Delta/2)^2-\Omega^2|b|^2}~.
\end{equation}
It is also worth noting that as shown in Fig. 1, the evolution 
of a single molecule, coupled to a single atomic mode, is 
always described by an oscillating solution, similar to 
\eq{aoscsol} since Bose enhancement, depicted by the exponential gain of 
Eqs. (\ref{asol}) and (\ref{nt}) is a collective effect, 
analogous to lasing.
\begin{figure}
\begin{center}
\epsfig{file=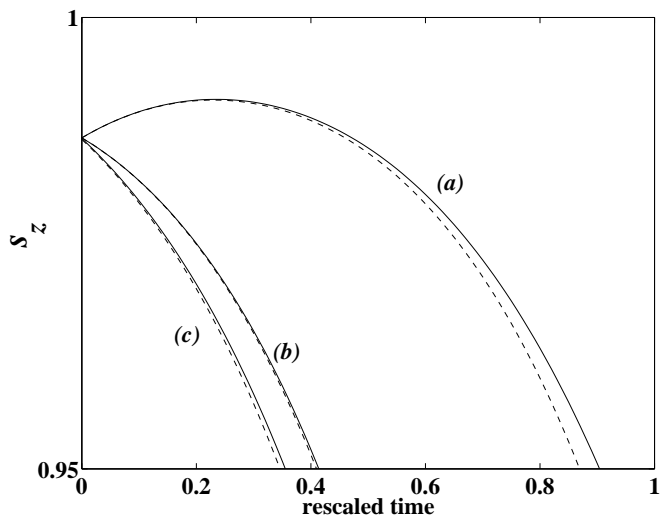,width=0.8\columnwidth,angle=90}
\end{center}
\caption{Depletion of the molecular mode for three 100 particle 
states with the same initial value of $\sz(0)=74/75$: 
(a) $(2/3)^{1/2}|0,N/2\rangle 
+ i(1/3)^{1/2}|2,N/2-1\rangle$,  
(b) $(2/3)^{1/2}|0,N/2\rangle - i(1/3)^{1/2}|2,N/2-1\rangle$, 
and (c) the state created by the spontaneous process starting from the pure 
molecular mode, according to \eq{nst} (- - -). Corresponding solid lines are 
'exact' numerical results. }
\label{f3}
\end{figure}  

Finally, we consider the dependence of the stimulated decay term
as described by \eq{nst}, on the exact initial conditions. Starting with 
an initially small coherent atomic amplitude [$|\langle \a(0)\a(0)
\rangle|=\langle \na(0)\rangle$] and assuming exact resonance ($\Delta=0$), 
the initial 
evolution of $n_{\text{st}}$ can be varied from exponential gain to 
exponential decay [$n_{\text{st}}(t)\sim \exp(\pm\lambda t)$]
 by controlling the relative phase between the  
atomic and molecular modes [$\langle \a(0)\a(0)\rangle$ and 
$b=\langle \b\rangle$]. Moreover, the squeezed state, formed by
spontaneous dissociation, always gains faster 
($\dot{n}_{\text{sp}}(t)/n_{\text{sp}}(t)=2\lambda \coth \lambda t >2\lambda$),
due to its high entanglement 
[$|\langle \a(t)\a(t)\rangle|/\langle \na(t)\rangle=\coth \lambda t >1$].  
In Fig. 3 we compare the predictions of 
\eq{nt} with exact quantum results for three initial amplitudes, 
corresponding to the same initial atomic occupation of $4/3N$.    
The (a) and (b) curves correspond to states of the form 
$c_0 |0,N/2\rangle+ic_2 |2,N/2-1\rangle$ with 
($c_0=\sqrt{2/3},c_2=i\sqrt{1/3}$) and 
($c_0=\sqrt{2/3},c_2=-i\sqrt{1/3}$), respectively, giving 
$\langle\b^*\langle\a(0)\a(0)\rangle=\pm i|b|\langle\na(0)\rangle$
(in equivalence to coherent states). 
The (c) lines depict the continued decay of a squeezed state with the 
same $\langle\na(0)\rangle$, formed by the purely spontaneous 
process, starting at $\tau=-0.53$. As expected, an initial 
exponential atomic loss is observed when $c_2=i\sqrt{1/3}$  
and an exponential atomic gain is observed when $c_2=-i\sqrt{1/3}$.
The gain of the spontaneously produced squeezed state, depicted 
by \eq{nsp}, is initially non-exponential,
becoming so only at later times. 

In conclusion, due to the dynamical instability of the molecular 
mode, the atomic ensemble produced by the dissociation stage of 
the parametric oscillation, is highly entangled. Consequently, 
atom-molecule coherent oscillations are predicted to be damped 
even in the two-mode limit. The leading quantum correction is 
an initially non-exponential spontaneous decay, becoming 
significant on a timescale which only grows as $\log{N}$. 
Stimulated processes taking place when there is an
initial atomic population, are sensitive to different initial 
conditions corresponding to the same $\sz(0)$. The 
fastest (super-exponential) decay rate is obtained for the initial 
state formed by spontaneous dissociation. This dependence originates 
in the parametric oscillation phase, driving it towards the molecular 
mode or away from it. It has significant implications on the
case of multiple atomic modes, suggesting that the rapid growth 
observed when these modes are initially partially populated 
\cite{goral} would be sensitive to the exact atomic amplitudes.

We are grateful to Vladimir Akulin for helpful discussions.
This work was supported by the National Science Foundation through
a grant for the Institute for Theoretical atomic and Molecular 
Physics at Harvard University and Smithsonian Astrophysical 
Observatory.

\end{document}